\documentclass{article}
\usepackage{spconf,amsmath,graphicx}

\usepackage[T1]{fontenc}
\usepackage[latin9]{inputenc}
\usepackage{float}
\usepackage{calc}
\usepackage{amsmath}
\usepackage{graphicx}
\usepackage{mathtools}
\usepackage[export]{adjustbox}
\usepackage{listings}
\usepackage{amssymb}
\usepackage{flexisym}
\usepackage{xcolor}
\usepackage{wrapfig}
\usepackage{textcase}
\usepackage{soul}
\usepackage{comment}

 \graphicspath{{./imgs/}}
 \usepackage[ruled,vlined]{algorithm2e}
\usepackage{tabularx,booktabs}
\usepackage[numbers,sort&compress]{natbib}
\usepackage{tikz}
\usetikzlibrary{positioning}

\title{Model-Based Reconstruction for Collimated Beam Ultrasound Systems}

\name{%
\begin{tabular}{@{}c@{}}
Abdulrahman Alanazi$^{1,3}$ \qquad 
Singanallur Venkatakrishnan$^{2}$ \qquad 
Hector Santos-Villalobos$^{2}$\\ 
Gregery Buzzard$^{1}$ \qquad 
Charles Bouman$^{1}$ 
\end{tabular}}
 \address{$^{1}$ Purdue University-Main Campus, West Lafayette, IN 47907.  \\
     $^{2}$ Oak Ridge National Laboratory, One Bethel Valley Road, Oak Ridge, TN 37831.\thanks{This manuscript has been authored by UT-Battelle, LLC under Contract No. DE-AC05-00OR22725 with the U.S. Department of Energy. The United States Government retains and the publisher, by accepting the article for publication, acknowledges that the United States Government retains a non- exclusive, paid-up, irrevocable, world-wide license to publish or reproduce the published form of this manuscript, or allow others to do so, for United States Government purposes. The Department of Energy will provide public access to these results of federally sponsored research in accordance with the DOE Public Access Plan (http://energy.gov/downloads/doe-public-access-plan).} \\
     $^{3}$King Saud University (KSU), Riyadh, Saudi Arabia.}

\begin{document}
\ninept
\maketitle

\begin{abstract}

Collimated beam ultrasound systems are a novel technology for imaging inside multi-layered structures such as geothermal wells. 
Such systems include a transmitter and multiple receivers to capture reflected signals.  
Common algorithms for ultrasound reconstruction use delay-and-sum (DAS) approaches; these have low computational complexity but produce inaccurate images in the presence of complex structures and specialized geometries such as collimated beams. 

In this paper, we propose a multi-layer, ultrasonic, model-based iterative reconstruction algorithm designed for collimated beam systems. 
We introduce a physics-based forward model to accurately account for the propagation of a collimated ultrasonic beam in multi-layer media and describe an efficient implementation using binary search.  
We model direct arrival signals, detector noise, and a spatially varying image prior, then cast the reconstruction as a maximum a posteriori estimation problem.
Using simulated and experimental data we obtain significantly fewer artifacts relative to DAS while running in near real time using commodity compute resources.

\end{abstract}

\section{Introduction}
\label{sec:intro}

Non-destructive characterization of multi-layered structures that can be accessed from only a single side is important for applications such as monitoring the structural integrity of oil and geothermal wells that lie behind layers of fluid and steel casing. 
A novel technology to address this problem is collimated beam ultrasound systems \cite{chillara2017low,chillara2018radial,chillara2019collimated,chillara2020ultrasonic}, which use carefully crafted acoustic beams with side lobes suppressed and transducer diffraction minimized to provide deep penetration and high spatial resolution.
Such a collimated beam source is combined with an array of receivers that capture the signals reflected by the target structure.  These signals are aggregated using a reconstruction algorithm to form an image of the structure.

Several classes of algorithms have been developed for general ultrasound reconstruction. 
The most popular of these use delay-and-sum (DAS) approaches for their  computational efficiency. 
One such approach is the synthetic aperture focusing technique (SAFT), which produces acceptable ultrasound images for simple objects \cite{prine1972synthetic}. 
SAFT has been applied to single-layer \cite{stepinski2007implementation,hoegh2015extended} and multi-layer \cite{skjelvareid2011synthetic,lin2018ultrasonic} structures but not to collimated beam systems. 
Multi-layer SAFT combines DAS with techniques such as ray-tracing \cite{cerveny2005seismic,shlivinski2007defect,margrave2019numerical} and root-mean-square velocity \cite{olofsson2010phase,lin2018ultrasonic} to compute the travel time in multi-layer media. 
However, SAFT and its variations rely on a simple model that often leads to artifacts such as multiple reflections. 
Furthermore, SAFT-based methods do not account for the shape of the beam or system noise, hence do not adapt easily to different acquisition geometries or noise levels. 
SAFT methods tend to lead to blurry images.  Methods to counteract this effect include \cite{jin2016sparse} and \cite{abdessalem2020resolution}, which use a linear forward model for single-layer structures and a carefully constructed sparse deconvolution approach. 

More physically realistic inversion methods include least-squares reverse time-migration (LSRTM) and full wave inversion (FWI) \cite{liu2016least,chen2018full}, both of which are used in seismology. 
LSRTM and FWI are iterative methods that seek the best least-squares fit between observed and reconstructed data based on a PDE model for wave propagation. 
These methods have the capability to image complex structures, but they rely on an iteration using a complex forward model, which makes them very computationally expensive. 

To reduce the reconstruction artifacts of SAFT while maintaining computational efficiency,  regularized inversion can be used with a linear propagation model.
In \cite{ozkan2017inverse}, the forward model is extended to handle plane-wave imaging.  
Most recently, the model-based iterative reconstruction (MBIR) approach of \cite{almansouri2018model} used a propagation model of the ultrasound through the media and combined all the data from the source-detector pairs to jointly reconstruct a fully 3D image.  
Nevertheless, no existing regularized MBIR method accounts for collimated beams and multilayered structures.

In this paper, we propose a multi-layer, ultrasonic, model-based iterative reconstruction (UMBIR) method for collimated beam systems. 
We introduce a new physics-based forward model to account for heterogeneous structures and introduce a binary search-based method to efficiently compute the travel time of acoustic waves in multi-layer media.
We use experimental measurements to modify the wave propagation model to account for collimated ultrasonic-beams, then  combine this physics-based model with a model for the direct arrival signals and noise in the detector to obtain a data-fidelity cost function.  We add a regularizing term using a spatially varying prior \cite{almansouri2018anisotropic} to cast the reconstruction as a maximum a posteriori estimation problem.
Finally, to maintain computational efficiency, we use the Iterative Coordinate Descent (ICD) algorithm to optimize our cost function.  Our results using simulated and experimental data demonstrate significantly fewer artifacts relative to DAS approaches while running in near real time using commodity compute resources.

\section{Multi-layer UMBIR}
\label{sec:main}
{\bf Notation:} 
Assuming a linear system for simplicity, we seek to reconstruct an image $x$ using a mathematical model of the form 
\begin{equation}
    \label{sysmodel}
    y = Ax + Dg + w, 
\end{equation}
where $y\in \mathbb{R}^{MK \times 1}$ is a vector of measurements from $K$ receivers at $M$ timepoints, $A \in \mathbb{R}^{MK\times N}$ is the system matrix, $x\in \mathbb{R}^{N \times 1}$ is the vectorized version of the desired image with $N$ total voxels, $D\in \mathbb{R}^{MK \times K}$ is a matrix whose columns form a basis for the possible direct arrival signals, $g \in \mathbb{R}^{K \times 1}$ is a scaling coefficient vector for $D$, and $w$ is a Gaussian random vector with distribution $N(0, \sigma^2 I)$. 
Conditioned on $x$ and $g$, $y$ is a Gaussian vector with the distribution $N(Ax + Dg,\sigma^2)$, so the negative log likelihood is 
\begin{equation}
    \label{eq:FW}
-\log p(y|x) = \frac{1}{2\sigma^2} \left\|  y-Ax-Dg\right\|_{2}^{2} + \text{constant}.
\end{equation}

{\bf Multi-layered structures:} Eq. \ref{eq:FW} transforms measurements into a prediction of their outcome image based on reflected signals due to heterogeneities in materials as well as direct arrival signals, which travel to the receivers directly with minimal interaction with the object.
In contrast to \cite{almansouri2018model}, we propose to modify the $A$ matrix to account for the specifics of the collimated beam systems which are the focus of this article. 
In order to compute entries of the $A$ matrix, we require an algorithm to compute the signal at each receiver for a given transmitted signal.  
\begin{figure}[htb]
\begin{center}
\includegraphics[width=0.35\textwidth]{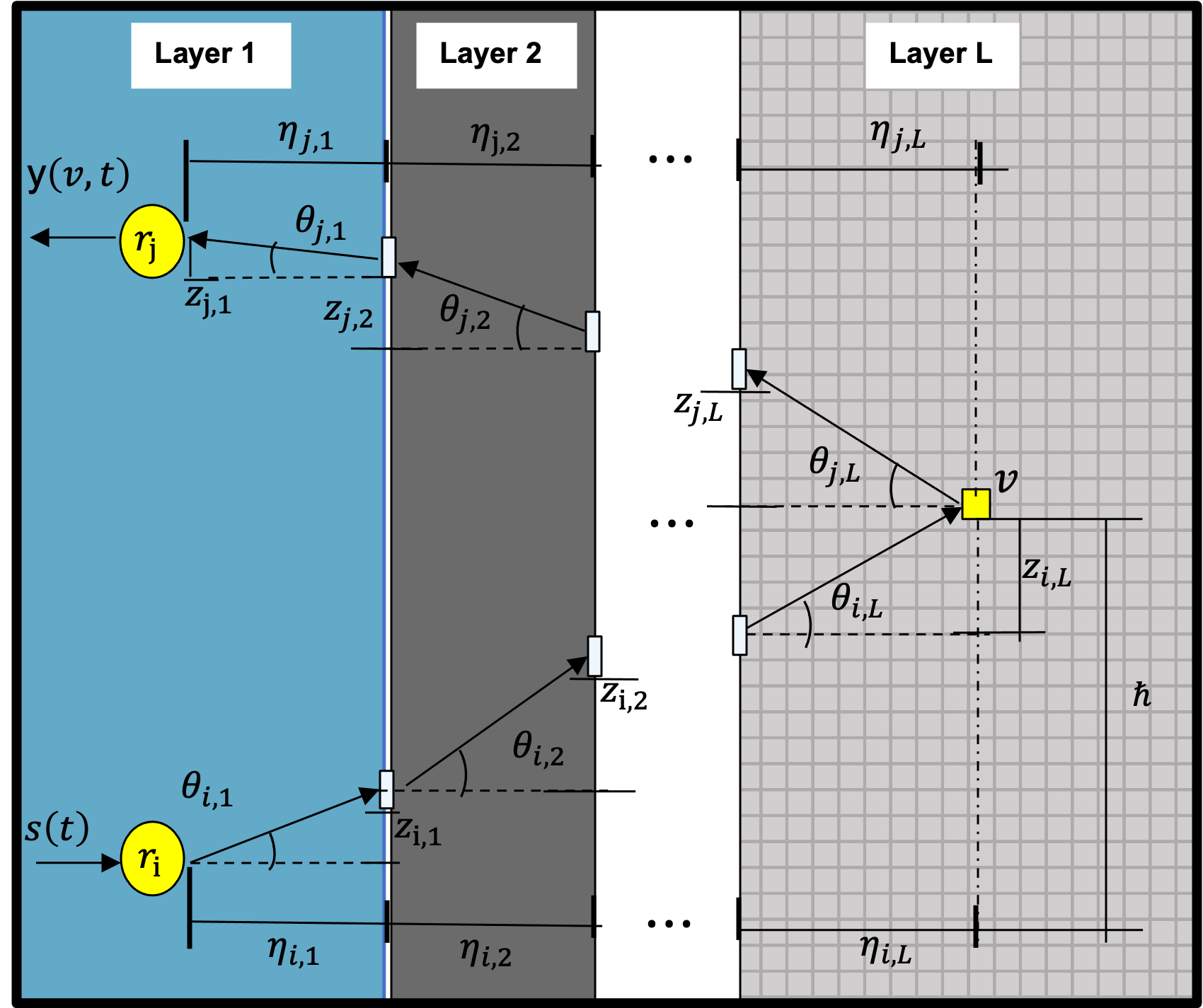}
    \flushleft\caption{Illustration of method used to compute time delays of ultrasound signals propagating through a multi-layers medium. 
    $s(t)$ is the input signal and $y(v,t)$ is the received signal from voxel $v$. We assume that we image the last layer.}
    \label{fig:MultiLay}
    \end{center}
\end{figure}
Following \cite{almansouri2018model} we assume that the frequency domain transfer function for a signal emitted at $r_i$, reflected at voxel $v$, and received at $r_j$ as in Figure~\ref{fig:MultiLay} is
\begin{equation}
    \label{eq:TransFerfunction1}
    G_{}(v,f) = \prod_{\ell=1}^{L} \tau_\ell e^{-(\gamma_{\ell}(v)|f|+2j\pi f T_{\ell}(v))}, 
\end{equation}
where $L$ is the total number of layers, $\tau_{\ell}$ is the transmittance coefficient of the front surface of the $\ell^{\text{th}}$ layer, $\gamma_{\ell}(v) = c_{\ell} \alpha_{\ell} T_{\ell}(v)$, $c_{\ell}$ is the acoustic speed in $\text{m}/\text{s}$ in the $\ell^{\text{th}}$ layer, $\alpha_{\ell}$ is the attenuation coefficient  in $\text{s}/\text{m}$ in the $\ell^{\text{th}}$ layer, and $T_{\ell}(v)$ is the travel time in seconds between the front and back interface of the $\ell^{\text{th}}$ layer (this depends on $i$ and $j$, which we suppress for notational clarity).

In single-layer structures, the computation of the time delays at image voxels is straightforward. 
However, in multilayered structures, the acoustic speed varies with depth, which causes reflections and refractions that result in a complex wave path. 
Analytically, the travel time from the source $r_i$ to the voxel $v$ to the receiver $r_j$ given the fixed depths $\eta_{i,\ell}$ and $\eta_{j,\ell}$, shown in Fig. \ref{fig:MultiLay}, can be computed according to Snell's law as 
\begin{equation}
    \label{eq:TD}
    T(v) = \sum_{\ell=1}^{L} \frac{\sqrt{z_{i,\ell}^2+\eta_{i,\ell}^2}+\sqrt{z_{j,\ell}^2+\eta_{j,\ell}^2}}{c_\ell},
\end{equation}
 where $z_{i,\ell} = \eta_{i,\ell} \text{tan}(\theta_{i,\ell})$ and $z_{j,\ell} = \eta_{j,\ell} \text{tan}(\theta_{j,\ell})$, $\ell = 1,2, \ldots, L$. 
 
 Note that there is no analytical solution to Eq. \ref{eq:TD} when the number of layers exceeds two. 
However, if we know the value of the incident angle $\theta_{i,1}$, finding the remaining unknowns is straightforward. 
To see this, consider for simplicity only the transmission path---the same steps can be used for the receiving path---from the source $r_i$ to the voxel $v$ shown in Fig. \ref{fig:MultiLay}. 
We assume that the thickness of each layer and the acoustic speed in each medium is known, in which case 
 the height of $v$ as a function of the $\theta_{i,\ell}$ is 
 \begin{align}
    \label{eq:h(v)}
    \sum_{\ell=1}^{L} z_{i,\ell} = 
    \eta_{i,1} \text{tan}(\theta_{i,1}) +
    \cdots +\eta_{i,L} \text{tan}(\theta_{i,L}). 
\end{align}
From Snell's law, we know that 
\begin{equation}
    \label{eq:thetas}
    \theta_{i,k} = \text{sin}^{-1}\left( \text{sin}(\theta_{i,k-1}) \frac{c_k}{c_{k-1}} \right), \, \forall k \in \{2, 3, \cdots, L\}. 
\end{equation}
From Eq.~\ref{eq:thetas}, each $\theta_{i,\ell}$ is an increasing function of $\theta_{i,1} \in [0, \pi/2]$, which implies that the height in Eq.~\ref{eq:h(v)} is an increasing function of $\theta_{i,1}$.  Hence we use binary search to find the value of $\theta_{i,1}$ so that Eq.~\ref{eq:h(v)} matches the height of $v$. 
Then Eq. \ref{eq:thetas} gives the corresponding angles and Eq.~\ref{eq:h(v)} gives the $z_{i,\ell}$.  The $z_{j,\ell}$ for the return path are found analogously, so Eq. \ref{eq:TD} gives the time delay $T(v)$.

Next, in frequency space, the received signal is proportional to 
\begin{equation}
    \label{eq:receivedSignal_freq}
    Y(v,f) = -x(v)S(f) \prod_{\ell=1}^{L} \tau_\ell e^{-(\gamma_{\ell}(v)|f|+2j\pi f T_{\ell}(v))},  
\end{equation}
where $x(v)$ in $\text{m}^{-3}$ is the reflection coefficient for the voxel $v$ and $S(f)$ is the Fourier transform of the transmitted signal. Define $\tau = \prod_{l=1}^L \tau_l$ and $\gamma (v) = \sum_{l=1}^L \gamma_l (v)$. In time-domain, the received signal is then 
\begin{equation}
    \label{eq:receivedSignal_time}
    y(v,t) = x(v) h(\gamma(v),t-T(v)),  
\end{equation}
where 
\vspace*{-\baselineskip}
\begin{equation}
    \label{eq:h}
    h(\gamma(v) ,t) = \mathcal{F}^{-1}\left\{ S(f) \tau e^{-\gamma(v) |f|} \right\}  
\end{equation}
and $\mathcal{F}^{-1}$ is the inverse Fourier transform. 
In order to reduce computation, we window in time and replace $h$ in Eq.~\ref{eq:receivedSignal_time} with
\begin{equation}
    \label{eq:h_apprx}
    \Tilde{h}(\gamma(v) ,t) = h(\gamma(v),t) \, \text{rect}\left( \frac{t}{t_0} - \frac{1}{2} \right),  
\end{equation}
$$
\text{rect}(u) = 1 \: \text{for} \: |u| < \frac{1}{2} \: \text{and}\: 0 \: \text{for} \: |u| \geq \frac{1}{2},
$$
\\
where $t_0$ is a constant based on the assumption that $h(\gamma(v),t)$ is equal to zero for $t > t_0$.
The signal received at time $t$ by transducer $r_j$ in response to the transmission from $r_i$ is computed by summing over all voxels $v$ to obtain
\begin{align}
    \label{eq:y_apprx}
    \Tilde{y}_{i,j}(t) &= \sum_v \Tilde{h}(\gamma(v),t-T(v))  x(v).
\end{align}
Summing over all transmitters $r_i$ and recalling that $T$ and $\Tilde{h}$ depend on $i$ and $j$, this linear relationship between $x(v)$ and $y(t)$ determines a single row of the system matrix $A$ in the time domain. 

\begin{figure}[htb]
 \centering
\includegraphics[width=0.4\textwidth]{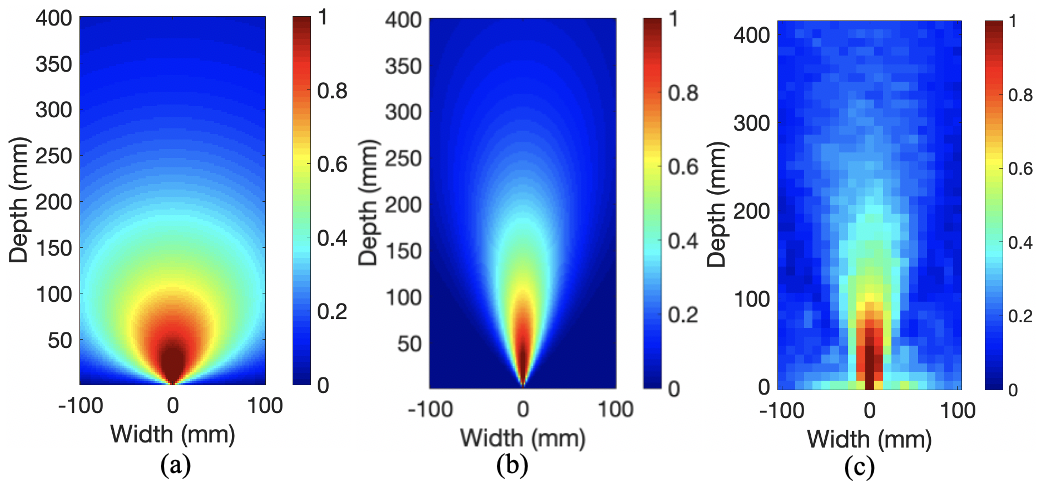} 
\caption{(a) A simulated beam profile, $\phi(v)^{(\beta)}$, with (a) $\beta=1$ and (b) $\beta=8$. (c) A real beam profile for a well-collimated source. 
$\beta$ can be tuned to match (b) to the experimental beam profile (c).}  
\label{fig:beams}
\end{figure}

{\bf Collimated beams:} 
In most reconstruction techniques, it is usually assumed that the wave propagation is isotropic; see \cite{tuysuzoglu2012sparsity}, for instance. 
However, this assumption is invalid for some systems and can produce artifacts in reconstructions \cite{chillara2020ultrasonic,chillara2019collimated}. 
Hence, we use a similar apodization function, $\phi(v)$, as in \cite{almansouri2018model} to reduce artifacts. 
However, we modify the apodization function to 
\begin{equation}
    \label{eq:apodFunc}
    \phi(v)^{(\beta)} = \text{cos}^\beta\left(\sum_{p=1}^{L} \theta_{i,p}\right) \, \text{cos}^\beta\left(\sum_{q=1}^{L} \theta_{j,q}\right),
\end{equation}
where the power $\beta$ controls the beam angular spread and can be tuned to achieve the desired beam collimation as illustrated in Fig. \ref{fig:beams}. 
Hence, we modify Eq. \ref{eq:y_apprx} to
\begin{equation}
    \label{eq:sysMatrix_timeMod}
    \Tilde{y}_{i,j}(t) = \sum_v \Tilde{h}(\gamma(v),t-T(v)) \phi(v)^{(\beta)} x(v)
\end{equation}
and again sum over $i$ to get the system matrix.
\vspace*{-\baselineskip}

{\bf Prior model:} For the prior model, we adopt the q-generalized Gaussian Markov random field (QGGMRF) from \cite{thibault2007three}.
With this design, the prior term is
\begin{eqnarray}
\label{eq:prior}
p(x) = \frac{1}{z}\exp\left(- \sum_{\{s,r\} \in C} b_{s,r} \ \rho(x_s-x_r)\right),
\end{eqnarray}
where $z$ is a normalizing constant, $C$ is the set of pair-wise cliques, 
\begin{eqnarray}
\rho(\Delta) = \frac{|\Delta|^p}{p\sigma_{g_{s,r}}^p}\left( \frac{|\frac{\Delta}{T\sigma_{g_{s,r}}}|^{q-p}}{1+|\frac{\Delta}{T\sigma_{g_{s,r}}}|^{q-p}}\right), \label{pot}
\end{eqnarray}
\vspace*{-\baselineskip}
\begin{eqnarray}
\sigma_{g_{s,r}} &=& \sigma_{0} \sqrt{m_s m_r}\label{SVR2},
\end{eqnarray}
\vspace*{-\baselineskip}
\vspace*{-\baselineskip}
\begin{eqnarray}
m_{s} &=& 1 + (m-1)*\left( \frac{\text{depth of pixel s}}{\text{maximum depth}}\right) ^a. \label{SVR}
\end{eqnarray}
Note that the condition $1<p<q=2$ must be satisfied to insure convexity and continuity of first and second derivatives of the prior model. 


{\bf Optimization:}
After designing the forward and prior models, we formulate our MAP estimate to be optimized. That is, by combining (\ref{eq:FW}) and (\ref{eq:prior}), we obtain 
\begin{eqnarray}
\begin{aligned}
(x,g)_{\text{MAP}} = \underset{x,g} {\text{arg\,min}} \biggl\lbrace &\frac{1}{2\sigma^2} \left\|  y-Ax-Dg\right\|^2\\
&+ \sum_{\{s,r\} \in C} b_{s,r} \ \rho(x_s-x_r) \biggl\rbrace.
\end{aligned} \label{final_map}
\end{eqnarray}
To optimize Eq. \ref{final_map}, we employ the Iterative Coordinate Descent (ICD) algorithm \cite{bouman2013model}. 
For convergence, it can be shown that our algorithm is guaranteed to converge to the global minimum since the cost function is continuously differentiable and strictly convex \cite{bouman2013model}. 
Once the forward and prior model are designed, one can find the pseudocode of UMBIR in \cite{almansouri2018anisotropic} for practical implementation. 
\begin{figure}[htb]
\centering
\includegraphics[width=0.45\textwidth]{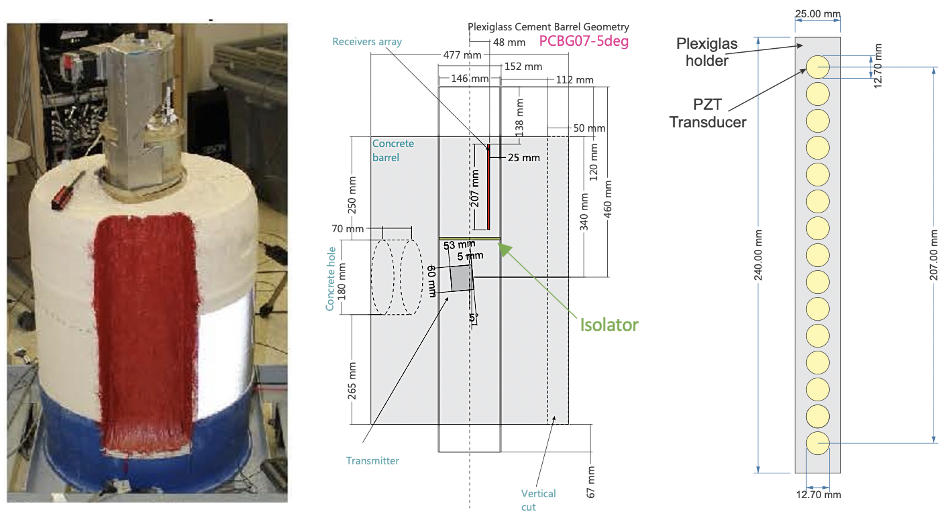}
\caption{(a) A picture of the specimen to be imaged using the collimated beam system.  The red region is a groove that is obtained by cutting a section of the cylinder. (b) The system geometry and (c) the receiver geometry.}
\label{fig:LANL_I_sys}
\end{figure}
\vspace*{-\baselineskip}
\vspace*{-\baselineskip}

\section{Experimental Results}
\label{sec:expr_results}
In this section, we compare UMBIR against SAFT using synthetic and real data sets. 
\begin{figure}[htb]
\centering
\begin{tabular}{cccc}
\includegraphics[width=2.5cm, height=4cm]{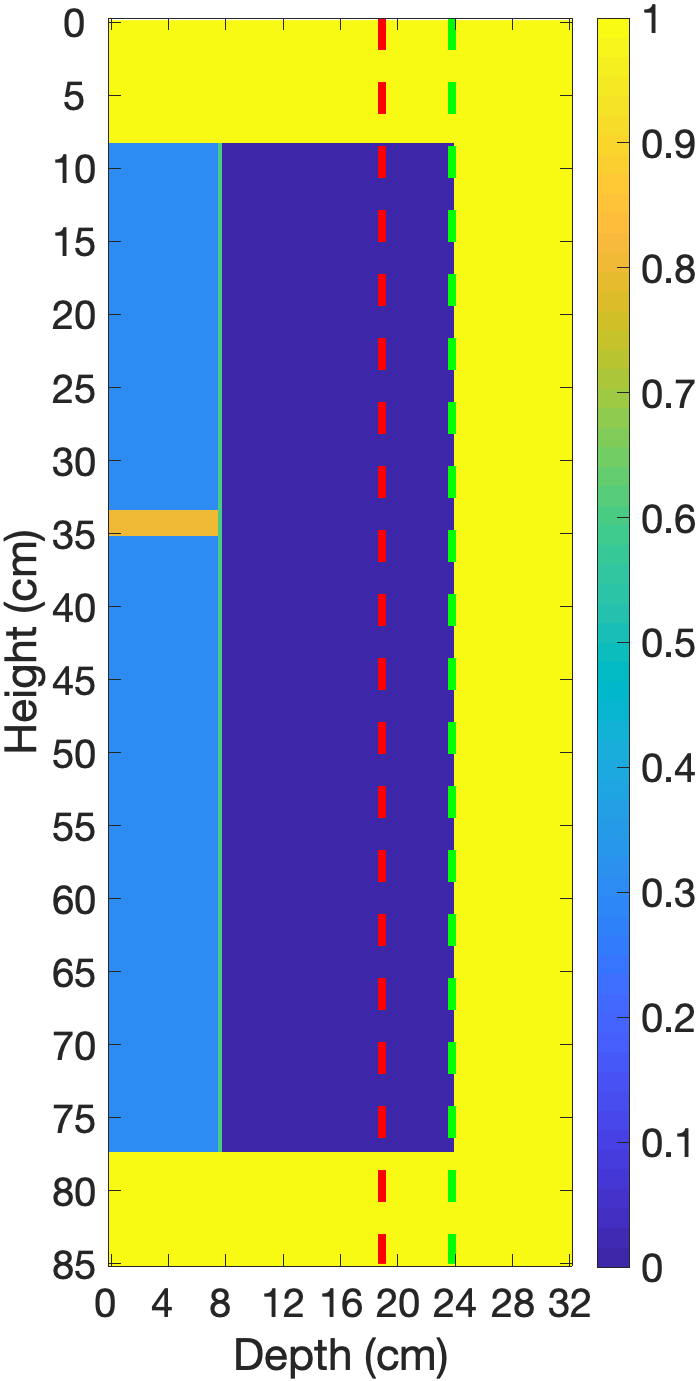} &
\includegraphics[width=2.5cm, height=4cm]{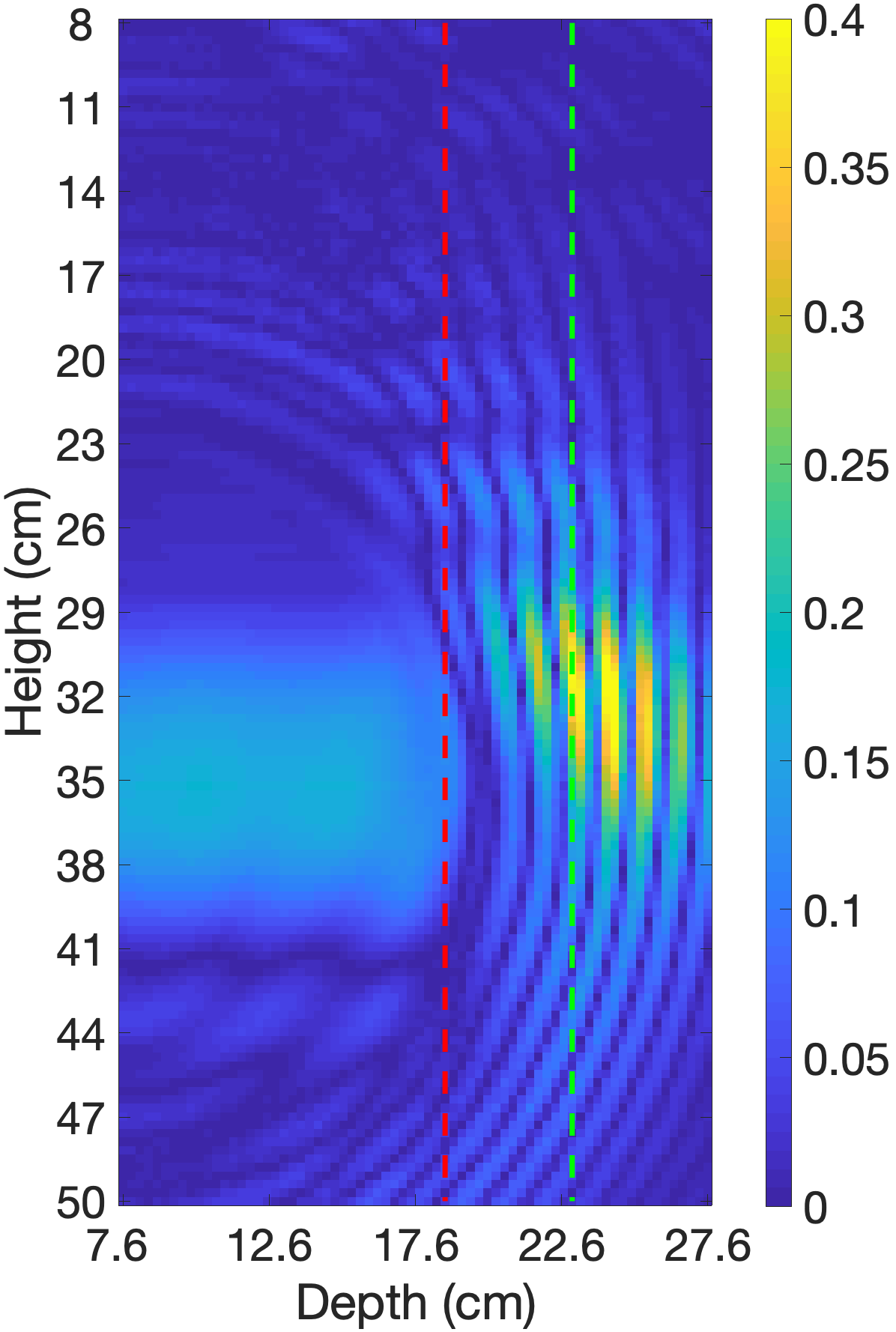} &
\includegraphics[width=2.5cm, height=4cm]{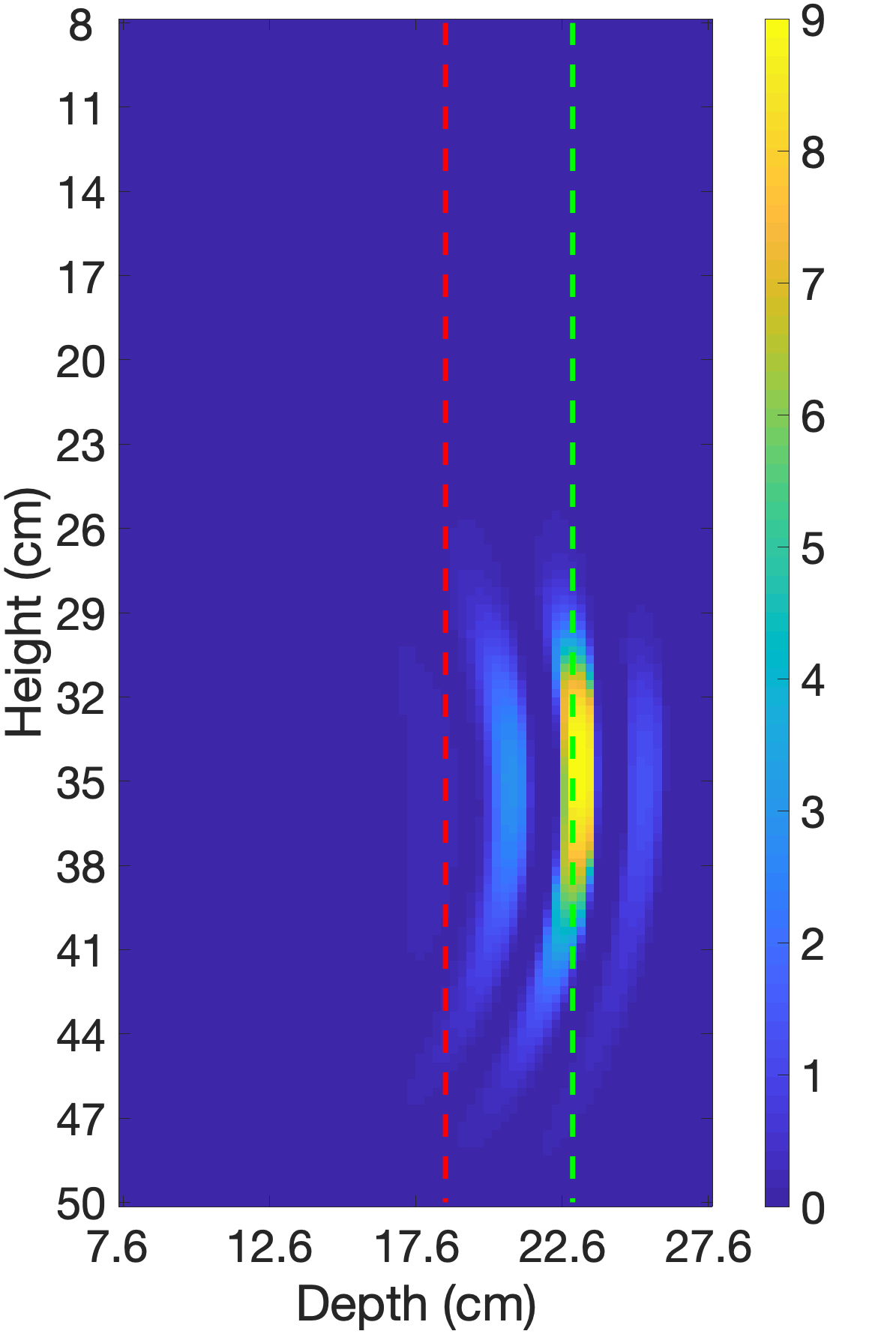}
 \tabularnewline
  \quad (a) K-Wave & (c) SAFT & (e) UMBIR
\tabularnewline
\includegraphics[width=2.5cm, height=4cm]{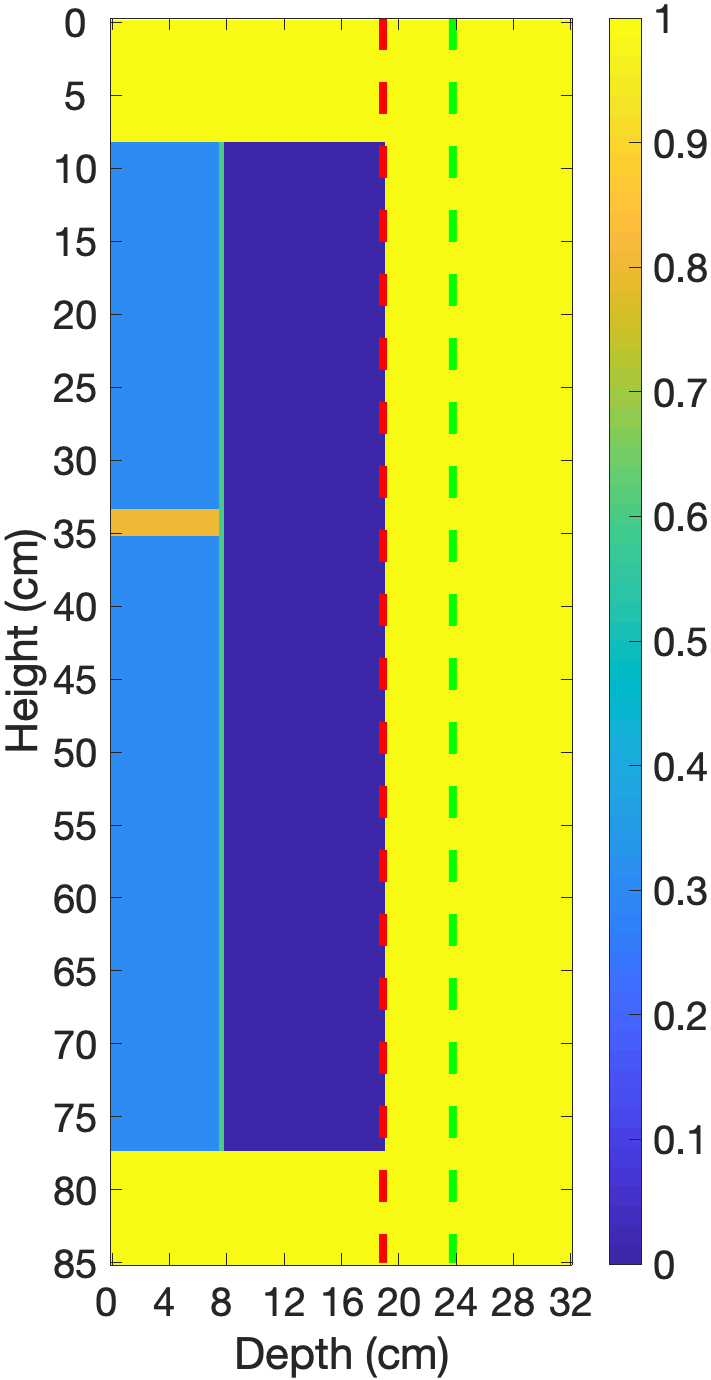} &
\includegraphics[width=2.5cm, height=4cm]{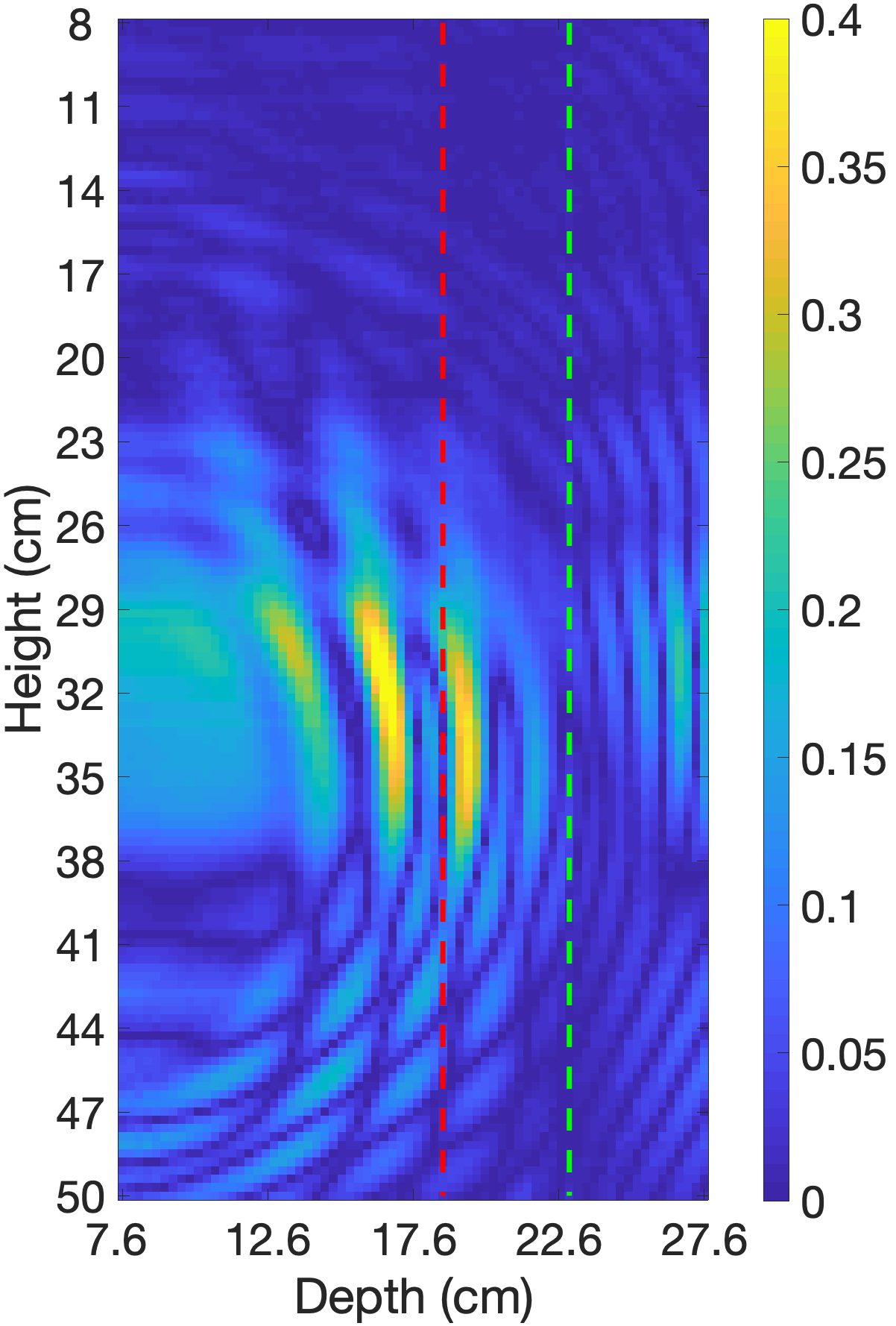} &
\includegraphics[width=2.5cm, height=4cm]{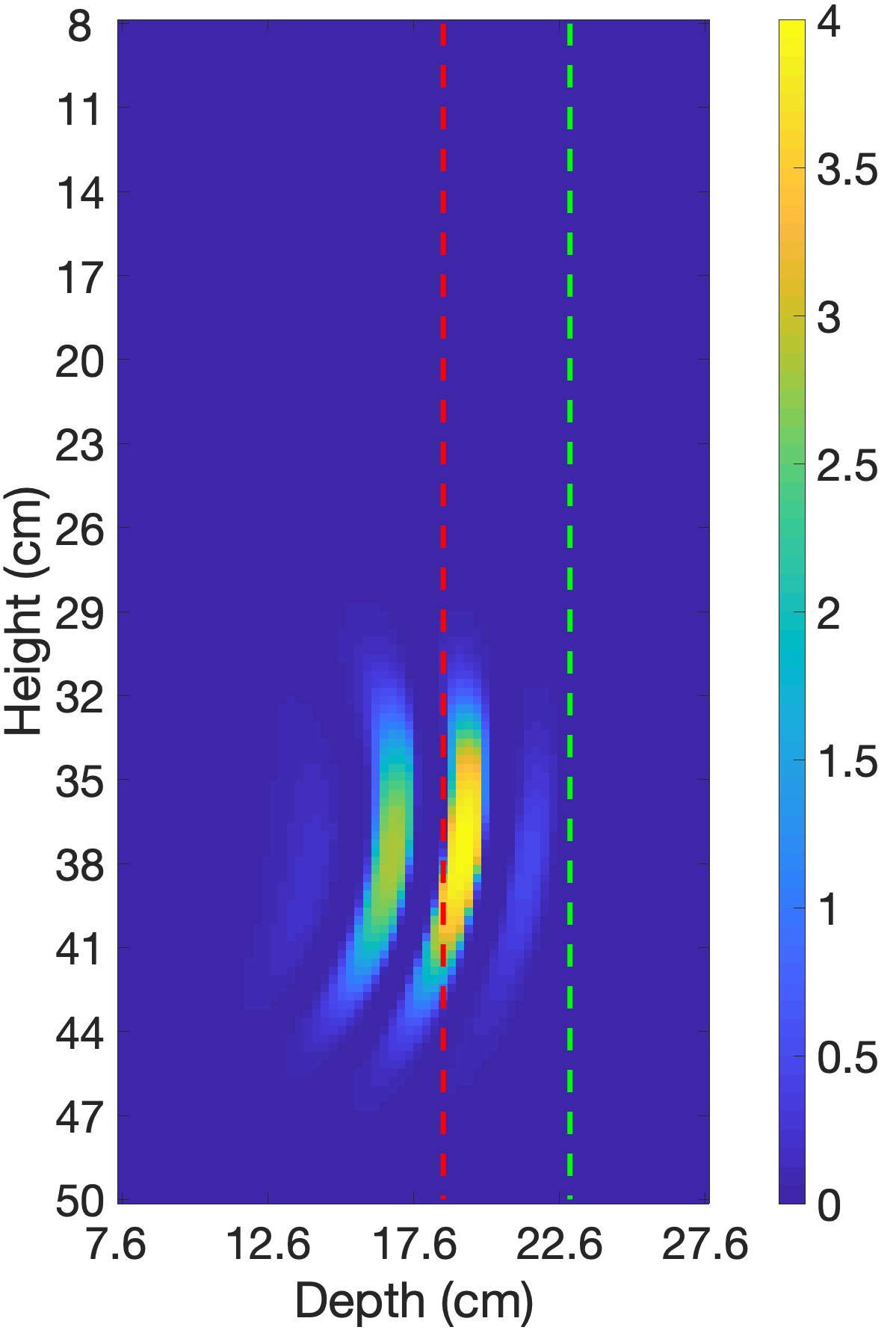}
 \tabularnewline
  \quad (b) K-Wave & (d) SAFT & (f) UMBIR
 \end{tabular}
\caption{(a) The ground truth used in K-Wave for the specimen without any defects and (b) with the groove. (c) and (d) SAFT reconstructions. (e) and (f) UMBIR reconstructions. The red and green dashed lines demonstrate the groove and backwall locations, respectively. UMBIR shows a clear improvement over SAFT.} 
\label{fig:LANL_I_kwaveResults}
\end{figure}
The experimental data was acquired from the concrete cylinder (CC) shown in Fig. \ref{fig:LANL_I_sys}(a) \cite{pantea2019collimated}. 
The sensor assembly is embedded in an  open borehole filled with water right in the center of the specimen. 
The open borehole is lined with a thin layer of Plexiglas that holds the entire system. 
The dimension of each layer is demonstrated in Fig. \ref{fig:LANL_I_sys}(b). 
There is only one intentional defect which is the groove (marked in red in Fig. \ref{fig:LANL_I_sys}(a)). 
The acoustic speed of the materials used in this experiment are 1.5 km/s, 2.82 km/s, and 2.62 km/s for the water, Plexiglas, and concrete layers, respectively. 
Also, the densities of the materials are 997 $\text{kg/m}^3$, 1180 $\text{kg/m}^3$, and 1970 $\text{kg/m}^3$ for the water, Plexiglas, and concrete layers, respectively. 
In this system, there is only one well-collimated beam transmitter (center of Fig. \ref{fig:LANL_I_sys}(b)) and 15 receivers mounted vertically (yellow discs in Fig. \ref{fig:LANL_I_sys}(c)). 
Furthermore, the data was collected for a rotational span of $180^\circ$, with a $5^\circ$ step size. 
Hence, the total number of scans is 37.
At the rotational position of $90^\circ$, the sensor assembly is facing the middle of the groove. 
In addition, before each run, the source was tilted upward by a firing angle of $5^\circ$. 
The input signal to the system has a central frequency of 58 kHz, and the data was sampled at 2 MHz. 
More details about the experiment can be found in \cite{pantea2019collimated,chen2018full}.
\begin{figure}[htb]
\centering
\begin{tabular}{cccc}
\tabularnewline
\includegraphics[width=6.5cm, height=2.8cm]{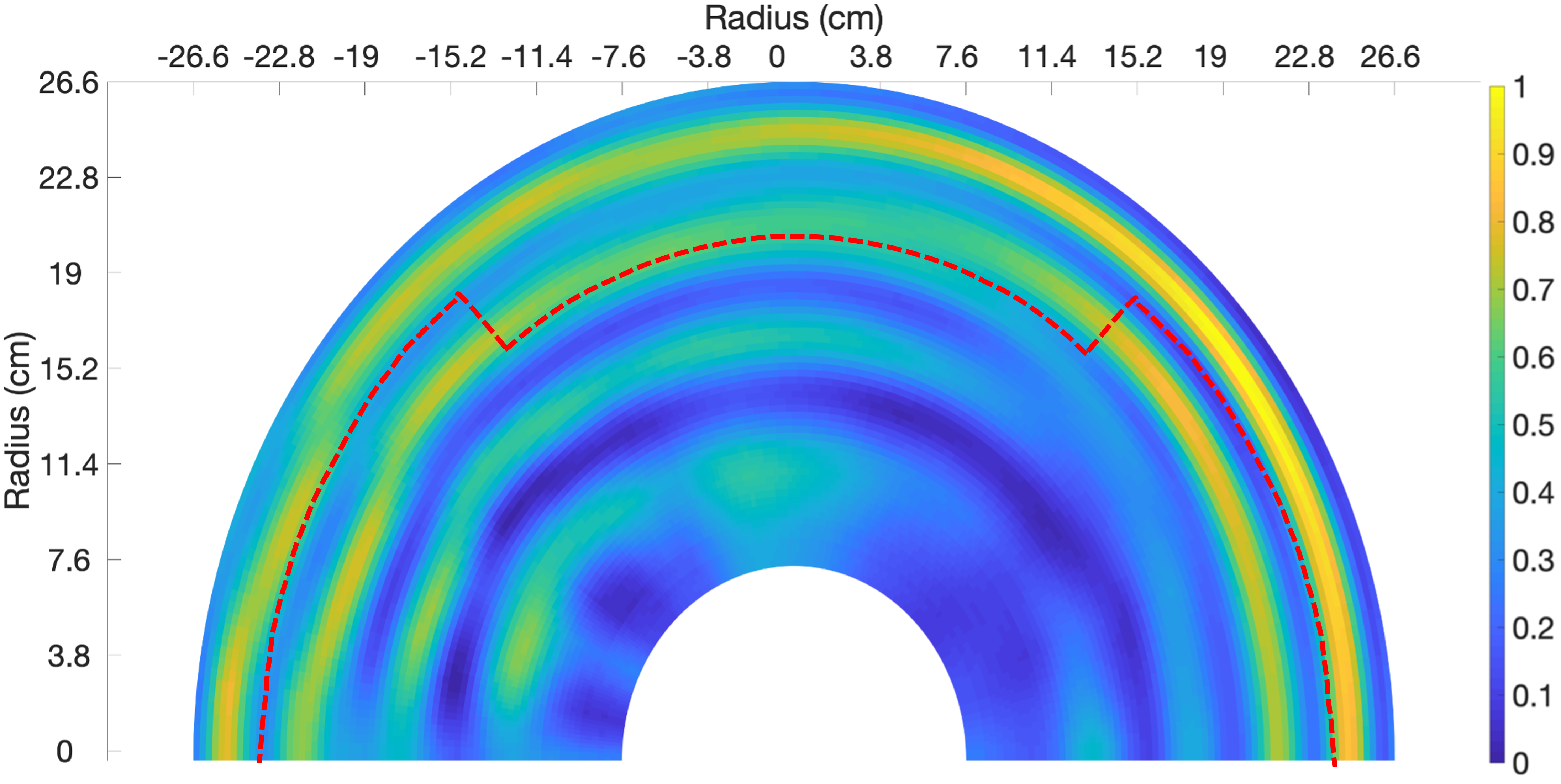} &
\includegraphics[width=1.3cm, height=2.5cm]{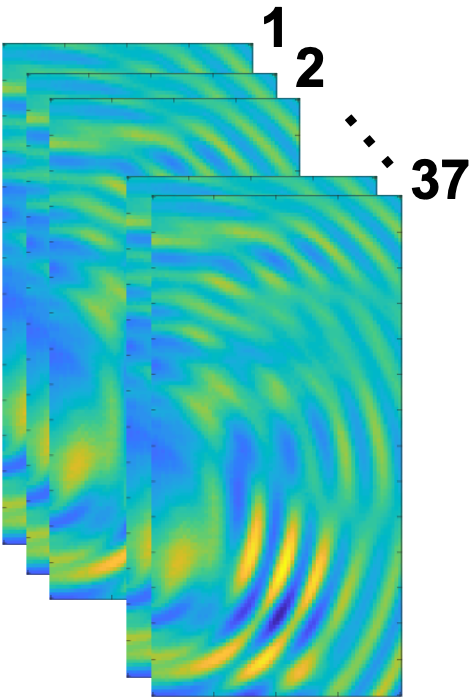} 
\tabularnewline
SAFT
\tabularnewline
\includegraphics[width=6.5cm, height=2.8cm]{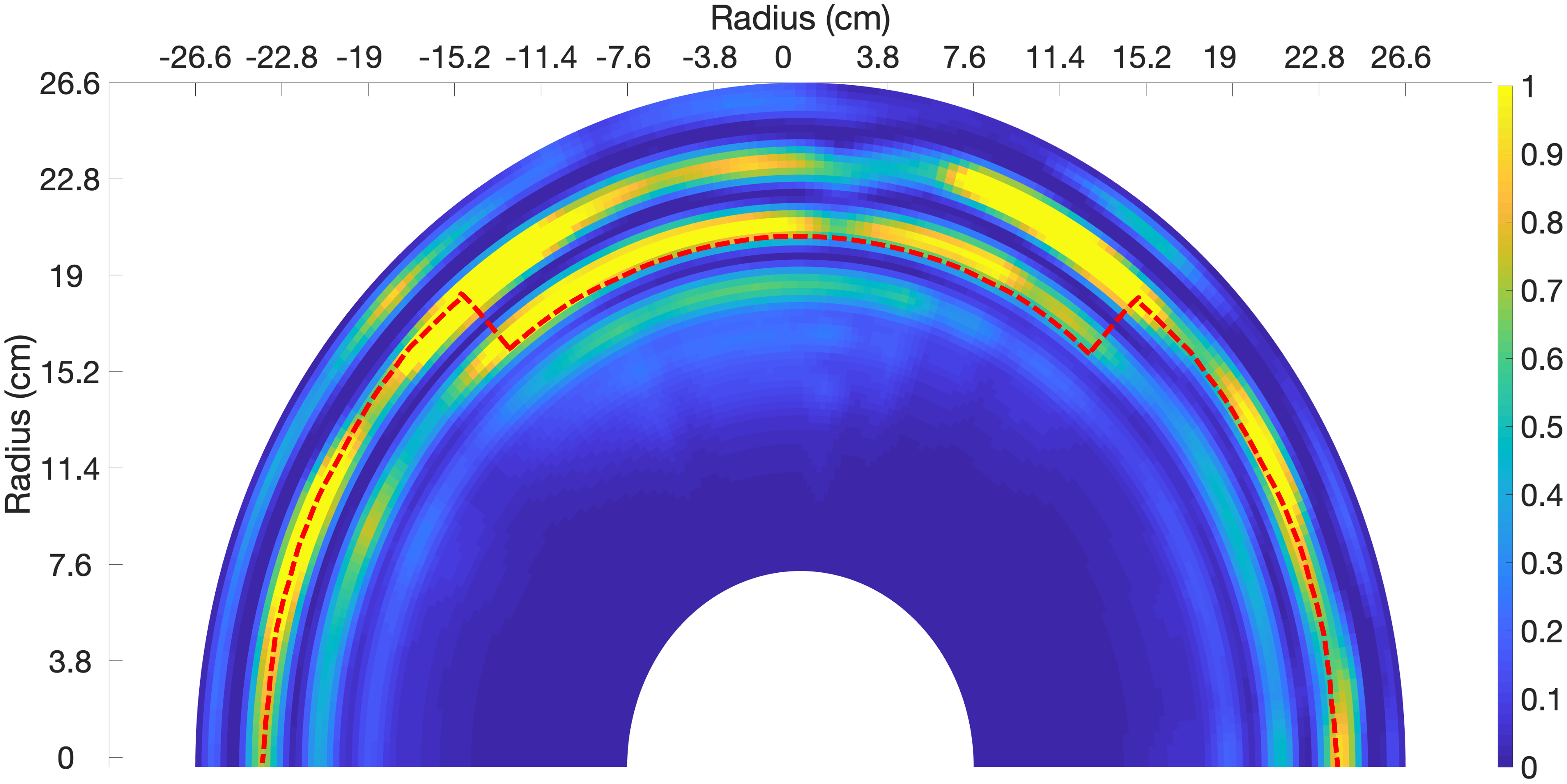} &
\includegraphics[width=1.3cm, height=2.5cm]{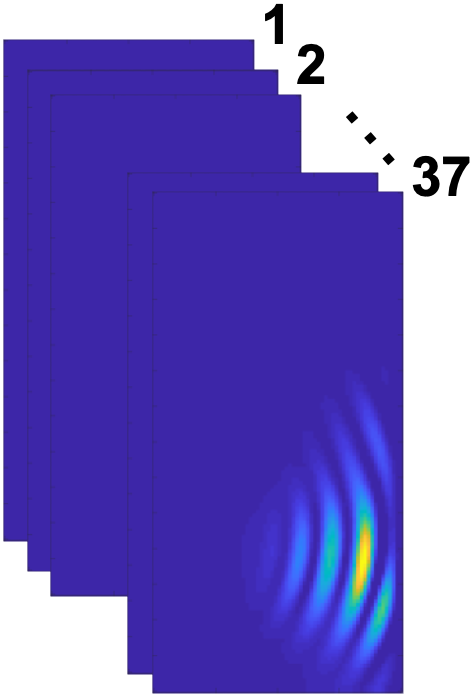} 
\tabularnewline
UMBIR
\tabularnewline
\end{tabular}
\caption{Left: Reconstruction of all views from real data using SAFT (top) and UMBIR (bottom). Right: Examples of cross-section reconstructions. The red dashed line specifies the outer boundary of the specimen. The groove and backwall are expected to be seen at radius of ~18 cm and ~23 cm, respectively.}
\label{fig:LANL_I_realResults_slices}
\end{figure}

Prior to testing our method on real data, we simulated two data sets for the CC experiment (with and without the groove) using 2D K-Wave simulations  \cite{treeby2010k}. 
The materials used along with their properties are set to the values used in the real experiment. 
The pixel pitch in both the horizontal and vertical axes is set to 1 mm. 
Since the computational domain in K-Wave has to be finite, it is required to use the perfectly matched layer (PML) \cite{treeby2010k}. 
In our simulations, we assume that the PML starts from the outer boundary of the computational domain and extends for 20 grid points in all directions. 
The source used in the simulations is a linear-array transmitter, tilted upward by $5^\circ$, and backed with an isolator to prevent the wave from propagating backward and be focused on the forward direction. 
The input signal used is the 58 kHz signal form the experimental data. 
All results in this work were performed on Intel(R) Core(TM) i7 CPU E5-2603 0 @1.80 GHz, 32.00 GB RAM. 


The UMBIR forward model parameters for reconstructing both simulated and real data were set to the following: $\beta = 8$, $\alpha_{\text{water}} = 2/\text{Hz m}$, $\alpha_{\text{concrete}} = 30/\text{Hz m}$, and the reconstruction resolution = 3 mm. 
When reconstructing the simulated data, the prior model parameters are: $\sigma = 0.01$ Pascal, $p=1.1$, $q=2$, $T=0.01$, $\sigma_0 = 5 \text{m}^{-3}$, $m=0.1$, and $a=1$. 
For the real data, the same previous prior model parameters used except that $\sigma =0.1$ Pascal and $m =0.13$. 
UMBIR stops after 50 iterations for K-Wave results and 100 iterations for the real data results.

Fig. \ref{fig:LANL_I_kwaveResults}(a) and (b) show the ground truths used in K-Wave to generate the data. 
The light blue region is water, dark blue is concrete, green is Plexiglas, and orange is a vacuum to block direct arrival signals. 
The yellow regions beyond the boundaries are air to mimic the real data.
SAFT reconstructions shown in Fig. \ref{fig:LANL_I_kwaveResults}(c) and (d) demonstrate reflections from the groove and backwall, respectively, with strong artifacts and reflection echoes. 
Conversely, the UMBIR reconstructions shown in Fig. \ref{fig:LANL_I_kwaveResults}(e) and (f) produce significantly better estimations of the groove and backwall and display far fewer spurious artifacts than SAFT. 
 
In order to visualize our results obtained from the real data for the 37 rotational positions, we implemented the following steps: 
First, we obtained a cross-section reconstruction for each of the 37 scans; examples of cross-section reconstructions are demonstrated in the right column of Fig. \ref{fig:LANL_I_realResults_slices}. 
Second, we selected a certain height in which reflections are visible. 
Finally, we interpolated the selected samples in polar coordinates by a factor of 5. 
Fig. \ref{fig:LANL_I_realResults_slices} shows SAFT (top) and UMBIR (bottom) reconstructions of the real data. 
Since the ground truth is unavailable, we validated our results with the imaging geometry shown in Fig. \ref{fig:LANL_I_sys}. 
According to Fig. \ref{fig:LANL_I_sys}(b), the groove and backwall are located at depths of 18.8 cm and 23.8 cm, respectively, from the borehole center. 
UMBIR reconstruction shows a clear reflection around a radius of 18 cm from groove. 
Also, at the rotational positions where the sensor assembly is not facing the groove, one can see a clear reflection from the backwall around 23 cm. 
In contrast, SAFT shows more and stronger artifacts than seen in UMBIR, and arguably displays more evidence of a groove where there is no groove than where there is a groove.

The computational complexity of our method is dominated by the complexity of computing the system matrix. 
In Section \ref{sec:main}, we facilitated the computation by repopulating the columns of the system matrix from $\Tilde{h}(.)$, which we need to compute only once. 
After constructing the system matrix, it can be used to perform reconstructions for all views. 
The UMBIR reconstructions shown above take about 30 to 40 seconds to construct the system matrix and about one minute to perform the reconstruction.

\vspace*{-\baselineskip}
\section{Conclusion}
 \vspace*{-\baselineskip}
\label{sec:conc}
In this paper, we proposed our multi-layer UMBIR algorithm designed for ultrasonic collimated beam systems. 
We showed the derivation of our modified forward model for multilayered structures and collimated ultrasonic-transducers. 
Our results demonstrated that our UMBIR shows clear improvements over SAFT and is effective for real data applications.

\vspace*{-\baselineskip}
\section{Acknowledgment}
\vspace*{-\baselineskip}
A. M. Alanazi was supported by King Saud University. 
C. A. Bouman was supported by the U.S. Department of Energy. 
G.T. Buzzard was partially supported by NSF CCF-1763896. 
S.V. and Hector Santos-Villalobos were supported by the U.S. Department of Energy staff office of the Under Secretary for Science and Energy under the Subsurface Technology and Engineering Research, Development, and Demonstration (SubTER) Crosscut program, and the office of Nuclear Energy under the Light Water Reactor Sustainability (LWRS) program.

\clearpage
\bibliographystyle{IEEEbib}
\bibliography{UMBIR_ICASSP2022}
\vfill\pagebreak

\end{document}